\documentstyle[hip-artc,epsfig,rotate]{article}  
\volnumber{5}  \edyear{1997}  \frompage{1} \topage{10}                
 \recrevdate{11 June 1997}                                              
\title{HBT with Space- vs. Time-like Hydrodynamic Freezeout} 
\authors{ 
{\twerm Henning Heiselberg
}\\[2.812mm]
{\normalsize
NORDITA, Blegdamsvej 17,\\ 
DK-2100 Copenhagen \O., Denmark\\[0.2ex] 
}}
 
\abstract{ 
Bose-Einstein correlations in relativistic heavy ion collisions 
and their dependence on the freeze-out condition in hydrodynamic models
is examined. The Cooper \& Frye {\it space-like} freeze-out mechanism
is compared to {\it time-like} freeze-out, where particles are emitted
{\it away} only from the surface, i.e. space- vs. time-like freeze-out. The
corresponding HBT radii are calculated for the two models emphasizing
the difference in the outward HBT radius. } 
 
\begin{document}
 
\maketitle

\section{Introduction}

Bose-Einstein interference of identical particles or the Hanbury-Brown
\& Twiss effect (HBT) \cite{HBT} shows up in correlation functions of
pions and kaons emitted from the collision zone in relativistic heavy
ion collisions. It is an important tool for determining the source at
freeze-out and recent data from relativistic heavy ion collisions can
restrict the rather different models, that have been developed to
describe particle emission in high energy nuclear collisions. In
hydrodynamical calculations particles freeze-out at a hypersurface
that generally does not move very much transversally until the very
end of the freeze-out \cite{Ruuskanen,Csernai,Schlei,Rischke}.  In
cascade codes the last interaction points are also found to be
distributed in transverse direction around a mean value that does not
change much with time \cite{Humanic,RQMD,QGSM,Pratt}, but the width of
the emission zone increases from narrow surface emission to a
widespread volume emission.  In the first stage the freeze-out surface
is relatively static in the transverse direction and the emission
extends in time whereas in the second stage the freeze-out happens
relatively rapid all over the spatial extent of the source.  The two
stages are often referred to as {\it surface} and {\it volume} freeze-out
respectively, or {\it time-like} and {\it space-like}
freeze-out (see Fig. 1).$^a$
The distinction is, of course, only approximate and the
amount of particles assigned to the two stages varies between models.

\begin{figure}[htb]
\vspace*{-1.0cm}
\centerline{\rotate[r]{\psfig{figure=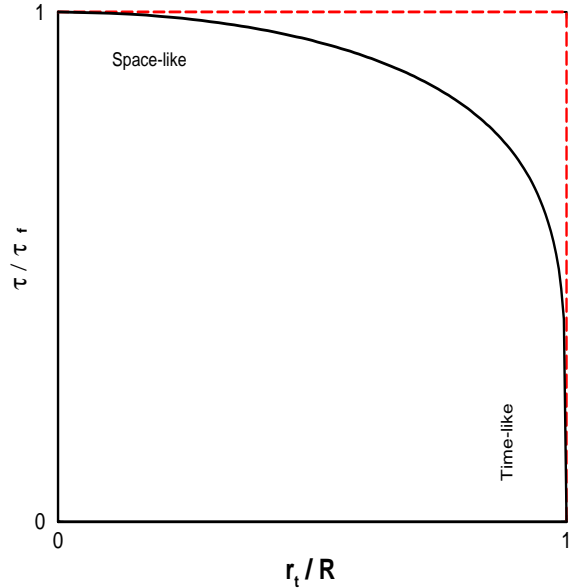,height=9cm,width=9cm}}}
\caption[]{Schematic picture of the transverse position and invariant time
of the freeze-out surface in hydrodynamic calculations. The dashed curve 
approximates the source by time- and space-like components.}
\label{fig1}
\end{figure}

In the Cooper \& Frye freeze-out assumption the net number of particles
leaving the hydrodynamic region is \cite{CF}
\begin{eqnarray}
  {\rm Cooper\,\&\, Frye\, (space-like) :}\quad\quad 
  E\frac{dN}{d^3p} = \frac{g}{2\pi} \int_{\sigma} d\sigma_\mu p^\mu f(x,p) \,,
\end{eqnarray}
where $f(x,p)=(exp(p\cdot u(x)/T(x)\pm 1)^{-1}$ is the standard thermal
Bose or Fermion distribution function with local flow velocity $u(x)$.
The freeze-out hypersurface is denoted by $\sigma_\mu$ and is usually 
defined at a constant energy density or temperature $T_c\simeq 100-150$ MeV
which is determined from the $p_\perp$ slopes.

It was pointed by Sinyukov \cite{Sinyukov}
that the Cooper \& Frye freeze-out was
inconsistent for time-like freeze-out since particles might reenter
the hydrodynamic phase. He therefore reflected the momenta in the
local rest frame at the boundary for those particles that would have
reentered the fluid 
\begin{eqnarray}
  {\rm Sinyukov\,(time-like):}\quad\quad 
  E\frac{dN}{d^3p} = \frac{g}{2\pi} \int_{\sigma} d\sigma_\mu p^\mu 
  f(x,{\cal R}_\sigma[p]) \,,
\end{eqnarray}
where ${\cal R}_\sigma[p]$ reflects the component of the particle 
momentum normal 
to the freeze-out surface in the local rest frame of the fluid.
The argument for doing this was that for each particle
reentering there is an outgoing particle inside the fluid that would have
escaped into the free streaming region. 
However, this is not true when the surface is moving (with velocity
${\bf v}_s$) in which case
slow particles may be overtaken by the freeze-out surface. 
Defining ${\bf n}_\sigma(x)$ as a vector normal to the freeze-out surface and 
directed away from the fluid and outward into the free streaming phase,
the condition for freeze-out is that particle velocities  
${\bf v_p}={\bf p}/E$ obey
\begin{eqnarray}
   ({\bf v_p} - {\bf v}_s(x))\cdot {\bf n}_\sigma(x) > 0  \,. \label{v}
\end{eqnarray}
Particles in the fluid cell to be frozen out 
that do not fulfill this condition should
therefore have their momenta reflected in the frame of the surface.
When the surface moves rapidly inwards not many particles need to have
their momenta reflected. Likewise, when there is strong outward directed
flow the thermal factor $\exp(p\cdot u(x))$ in distribution function 
automatically guarantees
that not many particles move slower than the surface speed.
But for small flow and slowly moving surfaces half of the particles
should be reflected.

The above prescription for correcting the time-like part of the Cooper
\& Frye freeze-out assumption may be considered as an
improvement. Yet, there are a number of considerations that remain
uncured.  It should be pointed out that this modified time-like
freeze-out assumption preserves momentum and energy globally in the
fluid, when it is applied to both sides, but not locally near the
surface. At the surface there is a net back reaction due to the
surface emission/evaporation that in the time-like freeze-out is {\it
away} from the surface only.  In the treatment of Godunov (see, e.g.,
\cite{Godunov}) restricting freeze-out to those particle for which the
four-vector product, $n(x)\cdot p>0$, is positive does, however, not
conserve global energy.  Also rescatterings and final state
interactions of emitted particles are still ignored.  Another issue is
the assumption of a sharp freeze-out surface. In reality the
freeze-out take place when the particle mean free path becomes long
enough that it can escape and the hydrodynamic assumption of a short
mean free path breaks down near the surface.  This is naturally
included in cascade models which find an extended region of final
interactions points around what is thus a diffuse surface
layer. Recently, Grassi et al. have improved the sharp surface
assumption in hydrodynamic models by a Glauber model approach
\cite{Grassi} (see also Csernai, these proceedings\cite{Csernai}).  
Similar ideas
have been discussed in \cite{Opaque} in connection with opaque sources
emitting from a surface layer of thickness $\sim \lambda_{mfp}$ and
their effects on HBT have been calculated.

\section{Space-like vs Time-like emission}

 We will discuss space-time correlations from hydrodynamic sources
with the Cooper \& Frye (space-like) and Sinyukov (time-like)
freeze-out assumption and compare them. In the subsequent chapter
we will then consider two-component sources with both space- and time-like
emission.

To study the differences between the freeze-out conditions we
restrict ourselves to a class of models that have:

{\it i) cylindrical symmetry around the beam or 
{\rm z}-axis},$^b$

{\it ii) longitudinal expansion with Bjorken scaling,}

{\it iii) no transverse flow.}$^c$

 The distribution
of emitted particles also referred to as the source of emission points
is with the Cooper \& Frye freeze-out assumption
\begin{eqnarray}
 S_{CF}(x,p) \sim e^{-p\cdot u(x)/T(x)}\, S_\tau(\tau) \delta(r_\perp-R(\tau))
   \,, \label{SCF}
\end{eqnarray}
Here, the Bjorken variable $\tau=\sqrt{t^2-{\rm z}^2}$ is the invariant
time and $\eta= 0.5\ln(t+{\rm z})/(t-{\rm z})$ the space-time rapidity.
The flow four-vector is $u=(\cosh(\eta),\sinh(\eta),0,0)$,
which gives $p\cdot u = m_\perp\cosh(\eta-Y)$.
$R(\tau)$ is the transverse radius of the freeze-out
surface that moves in time. Initially, when the nuclei collide
$R(0)$ is the transverse
size of the nuclear overlap zone and final freeze-out takes place
when $R(\tau_f)=0$. The surface speed is $v_s=dR(\tau)/d\tau$.
The temporal source factor $S_\tau(\tau)$ determines the amount of
particles that freeze-out per surface element at invariant time $\tau$.
Notice that any normalization is irrelevant as they cancel out in 
later correlation functions (\ref{C}) and HBT radii.

The Sinyukov freeze-out assumption for a time-like source without flow 
differ from the Cooper \& Frye by
\begin{eqnarray}
  S_{S}(x,p) = 2\,S_{CF}(x,p)\, \Theta({\bf p\cdot n}_\sigma(x)) 
             = 2\,S_{CF}(x,p)\, \Theta(\cos\theta) \,, \label{SS}
\end{eqnarray}
where $\theta$ is the polar angle between ${\bf p}$ and the vector
${\bf n}_\sigma(x)$ normal to the freeze-out surface and directed
away from the fluid and outward into the free streaming phase.
Notice, that the reflection does not change the single particle
distribution functions but only correlation functions that are
sensitive to correlations between $p$ and $x$.

\section{HBT Radii}

We follow the standard definition of correlation functions 
and the HBT radii as is described in more detail in the appendix.
It is common to boost longitudinally so that the pair rapidity always is
$Y=0$ and choose the direction of ${\bf p}=({\bf p}_1+{\bf p}_2)/2$ 
along the outward or {\rm x}
direction whereas the third {\rm y}-direction is called the sideward
directions. In cylindrical coordinates we choose the polar angle with respect
to the {\rm x}-axis and so the factor distinguishing the
Cooper \& Frye from the Sinyukov freeze-out is simply $\Theta(\cos\theta)$.

The HBT radii are for a space-like source $S_{CF}(p,x)$, 
for which $\langle{\rm x}\rangle=\langle\cos\theta\rangle=0$, 
\begin{eqnarray}
   R_l^2 &\equiv& \langle({\rm z}-\beta_l t)^2\rangle 
          \simeq  \langle\tau^2\rangle \frac{T}{m_\perp}  \,. \label{Rl}\\
   R_s^2 &\equiv& \langle{\rm y}^2\rangle 
         = \frac{1}{2} \langle R(\tau)^2\rangle           \,,\label{Rs}\\
   R_o^2 &\equiv& \langle({\rm x}-\beta_o t)^2\rangle 
         = \frac{1}{2}\langle R(\tau)^2\rangle 
          \,+\, \beta_o^2 \sigma(\tau)                    \,. \label{Ro}
\end{eqnarray}
The temporal and angular averages and fluctuations are described
in the appendix.

The outward HBT radius is thus larger than the sideward
\cite{Csorgo,Heinz}
\begin{eqnarray}
   R_o^2=R_s^2+\beta_o^2\sigma(\tau) \, . \label{R1}
\end{eqnarray} 
The measured out- and sideward
HBT radii are very similar in relativistic heavy ion collisions and is
based on Eq. (\ref{R1}) taken as an indication of very small duration
of emission, i.e., particles appear or freeze-out in a ``flash'' \cite{CC}.

Changing the Cooper \& Frye freeze-out to that of Sinyukov does not
change the sideward and longitudinal HBT radii since the reflection is
only in the outward or {\rm x}-direction.  However, the outward HBT
radius differ because the reflection has the effect that only
particles from the front side of the source pointing towards a given
detector are measured in that detector whereas those from the back
side are not.  The geometry is chosen such that the outward or {\rm
x}-axis is always along the particle momenta ${\bf p}\simeq{\bf p}$
which guaranties that only the half sphere toward any detector is seen
in that detector. This has the important consequence that $\langle{\rm
x}\rangle$ no longer vanishes but almost cancels $\langle{\rm
x}^2\rangle^{1/2}$. We find$^d$
\begin{eqnarray}
   R_o^2 &\equiv& (\frac{1}{2}-\frac{4}{\pi^2}) \langle R(\tau)\rangle^2
            +\beta_o^2\sigma(\tau) +\frac{1}{2}\sigma(R(\tau)) \nonumber\\
         &&  \,-\, 2\beta_o  \frac{2}{\pi}
             \langle(R(\tau)-\langle R(\tau)\rangle) 
            (\tau-\langle\tau\rangle)\rangle    \,. \label{RoS}
\end{eqnarray}
For a static surface $R(\tau)=constant$ the last two terms vanish;
for an inward moving surface the last term is positive.
The first term in (\ref{RoS}) is $\sigma({\rm x})$ and is significantly
reduced as compared to (\ref{Ro}).
As a consequence Eq. (\ref{R1}) no longer holds as was also found
for opaque sources or sources with strong transverse flow \cite{Opaque}. 

\section{Two Component Source}

It is important to distinguish between space- and time-like
freeze--out, since they give very different HBT radii.  Hydrodynamical
models~\cite{Csernai,Ruuskanen,Schlei,Rischke} assume that particles
are emitted from the surface. This is actually also found in some
cascade models at early times of the
collision~\cite{Humanic,RQMD,QGSM,Pratt}, but eventually the whole
source freezes out and disintegrates. The late stage of cascade models
resembles more a volume freeze--out.  These sources can approximately
be described by two components, initially surface emission but
eventually volume freeze-out. Generally, for a two-component source
$S(x)=pS_1(x)+ (1-p)S_2(x)$, properly normalized ($\int d^4x
S_i(x)=1$) such that $p$ is the fraction of particles from source 1,
the fluctuations in a quantity ${\cal O}$ is from (\ref{O})
\begin{eqnarray}
  \sigma({\cal O}) = p\,\sigma_1({\cal O}) + (1-p)\,\sigma_2({\cal O})
   +p\,(1-p)\,(\langle{\cal O}\rangle_1-\langle{\cal O}\rangle_2)^2 \,.
 \label{p}
\end{eqnarray}
Here, $\langle{\cal O}\rangle_i\equiv \int d^4x S_i(x){\cal O}$ and
$\sigma_i({\cal O})=\langle{\cal O}^2\rangle_i-\langle{\cal
O}\rangle_i^2$. The fluctuations are the weighted sum of the
fluctuations in the individual sources and an additional cross term.
Since $\langle {\rm y}\rangle=0$ and $\langle z-\beta_l t\rangle$ also
vanishes for $Y=0$ this additional cross term does not contribute to
the sideward and longitudinal HBT radii. It is, however,
nonnegligible for the outward HBT radius.

Let us approximate the qualitative features of hydrodynamic and
cascade models by a toy model with two-components: an approximately
static surface $R(\tau)=R$ for times $0<\tau<\tau_f$ and a rapid
freeze-out of the remaining source at $\tau=\tau_f$.  The first
component violates Eq. (\ref{v}) for the inward moving particles and
is thus time-like for which we should use the Sinyukov freeze-out
assumption.  The second component is space-like and we should apply
the Cooper \& Frye freeze-out assumption.  We weight the particles
frozen-out from the two component by the fractions $p$ and $(1-p)$
respectively.  The corresponding HBT radii are found by the
expressions (\ref{Rl}), (\ref{Rs}), (\ref{Ro}), and (\ref{RoS}) by
combining them according to (\ref{p})
\begin{eqnarray}
   R_l^2 &=&  (1-\frac{p}{2})\tau_f^2 \frac{T}{m_\perp}  \,. \label{Rlp}\\
   R_s^2 &=&  \frac{1}{4}(1+p) R^2        \,,\label{Rsp}\\
   R_o^2 &=&  \left( (\frac{1}{4}(1+p)
              -\frac{4}{\pi^2}p^2 \right)
              R^2 \,+\, \frac{p}{18}\beta_o^2 \tau_f^2    \,, \label{Rop}
\end{eqnarray}

\begin{figure}[htb]
\vspace*{-1.0cm}
\centerline{\rotate[r]{\psfig{figure=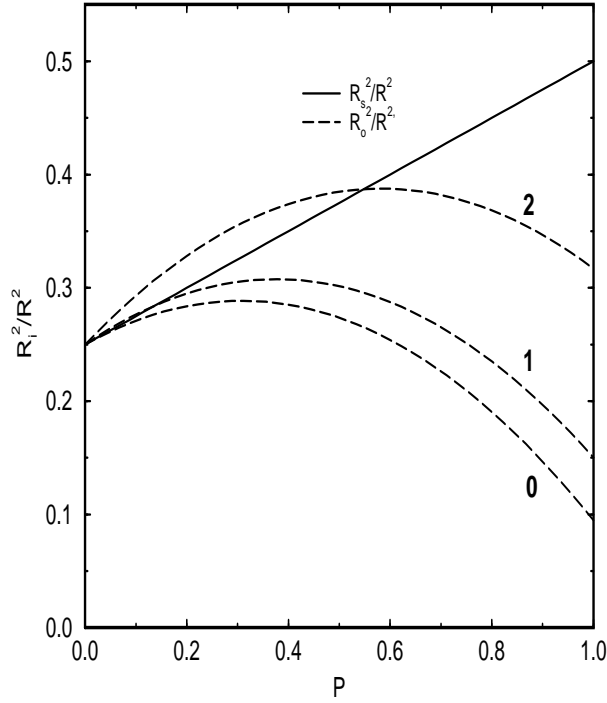,height=9cm,width=11cm}}}
\caption[]{Sideward and outward radii as function of the fraction of
time-like emitted particles $p$; the amount of space-like emitted
particles by the standard Cooper \& Frye freeze-out assumption is $(1-p)$. 
Dashed curves are $R_0^2/R^2$ with freeze-out times of 
$\beta_o\tau_f/R=0,1,2$. }
\label{fig2}
\end{figure}

In Fig. 2 we show the dependence of the out- and sideward HBT radii on
the space- vs. time-like fraction $p$.  The freeze-out assumption
clearly plays an important role and the difference $R_o^2-R_s^2$ is a
very sensitive quantity.  The fraction of time-like emission $p$ is a
model dependent parameter and in reality there will be a gradual
transition from time- to space-like emission. Also a finite width of
the emission layer \cite{Opaque} will diminish the difference.
Transverse flow has a similar effect of directed emission and leads to
further reducing $R_o$ with respect to $R_s$ \cite{Opaque}. However,
strong transverse flow $u_\perp$ also has the effect of improving the
condition of Eq. (\ref{v}) reducing range of particle velocities for
which time-like emission occur.

\section{Summary}

The validity of Cooper \& Frye space-time freeze-out
assumption has been discussed.  For sources with time-like emission it
breaks down and one should rather apply a modified freeze-out
assumption like that of Sinyukov which only allows emission away from
the fluid. It was shown in a simple toy model that the different
freeze-out conditions gave drastically different results for the
outward HBT radius.  A number of effects such as transverse
flow, diffuse surfaces, etc., reduce the difference between the two
freeze-out conditions.

In the hydrodynamic calculations of Ref. \cite{Rischke} at
RHIC energies the outward HBT radius $R_o$ is considerably larger than
$R_s$. This is mainly due to the existence of a long lived mixed phase
of quark-gluon plasma and hadronic matter for the equation of state
and initial conditions assumed in this calculation. Thus it is the
duration of emission that gives large $R_o^2-R_s^2$ or $R_o/R_s$ and
this part is unaffected by changing the freeze-out assumption from
Cooper \& Frye to that of Sinyukov as can be seen from Eqs. (\ref{Ro})
and (\ref{RoS}). At AGS and SPS energies, however, the measured $R_s$
and $R_s$ are very similar excluding the possibility for a long lived
mixed phase and the difference arising from the freeze-out
assumption may be significant.
\bigskip

\section*{Appendix: Correlation functions and HBT radii.}

For the correlation function analysis of Bose-Einstein interference
from a source of size $R$, we consider two particles emitted a distance
$\sim R$ apart with relative momentum ${\bf q}=({\bf p}_1-{\bf p}_2)$
and average momentum, ${\bf p}=({\bf p}_1+{\bf p}_2)/2$. Typical heavy
ion sources in nuclear collisions are of size $R\sim5$ fm, so that
interference occurs predominantly when 
$q\raisebox{-.5ex}{$\stackrel{<}{\sim}$}\hbar/R\sim 40$
MeV/c. Since typical particle momenta are $p_i\simeq p\sim 300$ MeV,
the interfering particles travel almost parallel, i.e.,
$p_1\simeq p_2\simeq p\gg q$.  The correlation function due to
Bose-Einstein interference of identical particles from an incoherent
source is (see, e.g., \cite{Heinz})
\begin{equation}
   C_2({\bf q},{\bf p})=1\;\pm\; |\frac{\int d^4x\;S(x,{\bf p})\;e^{iqx}}
   {\int d^4x\;S(x,{\bf p})}|^2 \,, \label{C}
\end{equation}
where $S(x,{\bf p})$ is a function describing the phase space density of the
emitting source. The $+/-$ refers to boson/fermions respectively.

Experimentally the correlation functions for identical mesons
($\pi^\pm\pi^\pm$, $K^\pm K^\pm$, etc.) are often
parametrized by the gaussian form
\begin{equation}
  C_2(q_s,q_o,q_l)=1+\lambda\exp(-q_s^2R_s^2-q_o^2R_o^2-q_l^2R_l^2
  -2q_oq_lR_{ol}^2  )\;.   \label{Cexp}
\end{equation}
Here, ${\bf q}={\bf k}_1-{\bf k}_2=(q_s,q_o,q_l)$ is the relative momentum
between the two particles and $R_i,i=s,o,l$ the corresponding sideward,
outward and longitudinal HBT radii respectively.
We will employ the standard geometry, where the {\it longitudinal} direction
is along the beam axis and the outward direction is along ${\bf p}$ and
the sideward axis is perpendicular to these.
Usually, each pair of mesons is lorentz boosted longitudinal to the system 
where their rapidity vanish, $Y=0$. Their average momentum ${\bf p}$ 
is then perpendicular to the beam axis and is chosen as the {\it outward}
direction. In this system the pair velocity
\mbox{\boldmath $\beta_p$}=${\bf p}/E_p$ points in
the outward direction with $\beta_o=p_\perp/m_\perp$ where 
$m_\perp=\sqrt{m^2+p_\perp^2}$ is the transverse mass.
As pointed out in \cite{Heinz} the out-longitudinal coupling $R_{ol}$
vanishes to leading order when $Y=0$.
The reduction factor $\lambda$ in Eq. (\ref{Cexp}) may be
due to long lived resonances \cite{Csorgo,HH}, coherence effects,
incorrect Gamov corrections or other effects. It is found to
be $\lambda\sim 0.5$ for pions and $\lambda\sim 0.9$ for kaons.

It is convenient to introduce the source average and fluctuation
or variance of a quantity ${\cal O}$ defined by
\begin{equation}
 \langle{\cal O}\rangle\equiv
   \frac{\int d^4x\;S(x,{\bf p}){\cal O}}{\int d^4x\;S(x,{\bf p})}\;, \quad
 \sigma({\cal O}) \equiv \langle{\cal O}^2\rangle -\langle{\cal O}\rangle^2
    \;. \label{O}
\end{equation} 
With $qx\simeq{\bf q\cdot x}-{\bf q\cdot}$\mbox{\boldmath $\beta$}$_{\bf p}\,t$
one can, by expanding to second order in $q_iR_i$ and comparing to
Eq. (\ref{Cexp}), find the HBT radii $R_i$, {\it i=s,o,l}.
They are \cite{Heinz}
\begin{equation}
   R_i^2=\sigma(x_i-\beta_i\;t) \,. \label{Ri}
\end{equation}
The HBT radii are a measure for the fluctuations of
$(x_i-\beta_it)$ over the source emission function $S$.

In the local center of mass system defined by $Y=0$ we have
$\beta_l=0$ and $\beta_o=p_\perp/m_\perp$. The HBT radii
reduce to (${\rm x}=R(\tau)\cos\theta$ and ${\rm y}=R(\tau)\sin\theta$)
\cite{Fluc}
\begin{eqnarray}
   R_l^2 &\equiv& \sigma({\rm z}) = \sigma(\tau\sinh\eta)
          \simeq  \langle\tau^2\rangle \frac{T}{m_\perp} 
          \,. \label{Rla}\\
   R_s^2 &\equiv& \sigma({\rm y}) 
         =  \langle R(\tau)^2\rangle \sigma(\sin\theta)
         =  \frac{1}{2}\langle R(\tau)^2\rangle \,,\label{Rsa}\\
   R_o^2 &\equiv& \sigma({\rm x}-\beta_o     t)
         = \langle R(\tau)\rangle^2 \sigma(\cos\theta)
             \,+\, \sigma(R(\tau))\langle\cos^2\theta\rangle
             \,+\, \beta_o^2 \sigma(\tau)\nonumber\\ 
         &&  \,-\, 2\beta_o    \langle\cos\theta\rangle
             \langle(R(\tau)-\langle R(\tau)\rangle)
            (\tau-\langle\tau\rangle)\rangle      \,. \label{Roa}
\end{eqnarray}

The averages simplify because 
the space-time rapidity, angular and temporal integrations separates and
due to the normalization a number of factors cancel. 
For example, a function of proper time only needs to be
averaged with respect to the temporal parts of the source 
\begin{eqnarray}
   \langle {\cal O}(\tau)\rangle=\frac{\int^{\tau_f}_0d\tau\,\tau R(\tau)\,
           S_\tau(\tau) {\cal O}(\tau)}
           {\int^{\tau_f}_0d\tau\,\tau R(\tau)S_\tau(\tau)} \,. \label{tau}
\end{eqnarray}
In the toy model employed above the emission per surface element 
$S_\tau(\tau)$ is assumed constant.

The angular averages also simplify for cylindrical geometry.
>From the definitions in Eq. (\ref{O}) we obtain
\begin{eqnarray}
 \langle {\cal O}(\theta)\rangle 
 &=& \frac{\int_{-\pi/2}^{\pi/2} {\cal O}(\theta) 
    \cos\theta\, d\theta }
   {\int_{-\pi/2}^{\pi/2} \cos\theta\, 
     d\theta }      \,. \label{Theta}
\end{eqnarray}  
Notice that always $\langle {\rm y}\rangle=\langle\sin\theta\rangle=0$,
whereas $\langle {\rm x}\rangle=\langle R(\tau)\rangle
\langle\cos\theta\rangle\ne 0$, when cylindrical symmetry is broken as
for the Sinyukov freeze-out condition, opaque sources, or 
sources with transverse flow. With transverse flow $u(r_\perp)$ the
thermal factor leads to a factor $\exp({\bf p}\cdot{\bf u}(r_\perp)/T)$
in Eq. (\ref{Theta}) \cite{Fluc}
which moves the measured emission region in the
direction towards ${\bf p}$ or the detector. 

\section*{Acknowledgement}
Discussion with L. Csernai, L. McLerran, D. Rischke,
V. Ruuskanen, R. Venugopalan,
and A. Vischer are gratefully acknowledged as well as the ECT$^*$
workshop in May 1997 on hydrodynamics.

\section*{Notes}
\begin{notes}
\item[a]
The terminology chosen here is to call surface
emission {\it time-like} as it takes place from a small volume but for
a long time, whereas volume emission is {\it space-like} as it occurs
in a large volume in a short period of time. The opposite terminology
is sometimes used when referring to the direction of the four-vector
$n_\sigma(x)$ normal to the freeze-out hypersurface.
\item[b]
However, cylindrical symmetry is broken by
the direction of the detector when, for example, the source
is opaque, has  transverse flow, or for time-like freeze-out.
\item[c]
Transverse flow has been 
included in the case of opaque sources \cite{Fluc}. We will comment on
the effect of flow later.
\item[d]
 In case of time-like freeze-out,
the factors $2/\pi$ 
arise from the nonvanishing angular average
$\langle\cos\theta\rangle$ $ =$ $ \int^{\pi/2}_{-\pi/2}\cos\theta\,d\theta/
\int^{\pi/2}_{-\pi/2}d\theta$ $ =$ $ 2/\pi$.
\end{notes}

\vfill\eject
\end{document}